\documentclass[aps,prd,onecolumn,groupedaddress,showpacs,nofootinbib,amssymb]{revtex4}
\usepackage[dvips]{graphicx}
\usepackage{amssymb}
\usepackage{amsmath}
\usepackage{amsfonts}
\usepackage{bm}
\usepackage{latexsym,float}
\usepackage{amssymb}
\usepackage{verbatim}
\usepackage{multirow}
\usepackage{color}

\begin{document}

\title{Spin connection and cosmological perturbations in scalar-torsion  gravity.}


\author{Alexey V. Toporensky$^{1,2}$}
\email{atopor@rambler.ru} \affiliation{$^{1)}$SAI MSU, Moscow, Russia\\ $^{2)}$Institute of Physics, Kazan Federal University,
Kremlevskaya street 18, 420008 Kazan, Russia}

\author{Petr V. Tretyakov$^{3}$}
\email{tpv@theor.jinr.ru} \affiliation{$^{3)}$Joint Institute for
Nuclear Research, Joliot-Curie 6, 141980 Dubna, Moscow region,
Russia}

\begin{abstract}
In this paper we generalize our previous results on spin connection for the linear scalar cosmological perturbation in $f(T)$ theory to wider class of theories which includes
a scalar field $\Phi$ non-minimally coupled to torsion, and allows  $\Phi$-dependence of the function $f$. The case of $f_{TT}=0$, with a scalar field non-minimally coupled to
torsion needs a special treatment. In both considered cases we derive self-consistent equations of motion for cosmological perturbation using a solution for non-trivial spin connection.

\end{abstract}


\maketitle


Recently new type of modified gravity theories based on generalizations of Teleparallel Equivalent of General Relativity (TEGR) became very 
popular topic of intense investigations. In TEGR it is the torsion $T$ which encodes the deviation from Minkowski space, while in classical formulation
of General Relativity (GR) this role is played by the curvature $R$ \cite{Einstein}. As curvature-free connection is not unique (in contrast to torsion-free Levi-Civita connections uniquely determined by metric), the additional structure in the form of tetrad field is
required for the correct formulation of the theory \cite{AP}. This tetrad field decouples from the resulting equations of motion.
The modification of TEGR is usually made by the same method as in modified gravity
based on GR, replacing the torsion scalar $T$ in the action by some function $f(T)$. It is known that despite TEGR is equivalent to GR at the level of equations of motion,
$f(T)$ and $f(R)$ theories with the same function $f$ are different theories \cite{Linder}. A nice feature of $f(T)$ theory is that its equations of motion are second order differential equations
in contrast to fourth order differential equations for a $f(R)$ theory. This fact may be considered as a significant advantage of $f(T)$ modifications of gravity \cite{BMNO,BCNO,RHSGR,BOSG,BCLNSG,N1,CCLS,N2,CLSX,BDERFGHHSMV}. However,
this class of theories has a serious problem already at the level of equations of motions  \cite{LSB,SLB}. In the first version of the theory
it appears that only a special tetrad (a "good tetrad") is required on order to get consistent equations of motion \cite{arg,TB}. Otherwise, 
equations requires that second derivative of the function $f$ with respect to $T$ (usually denoted as $f_{TT}(T)$) should vanish,
with simply means that we are back in TEGR \cite{CDDS}.

Later, new version of the theory have been invented, in this version it is possible to
use any tetrad if it is accompanied by suitable spin connection \cite{KS2,G2,HJU,G3,KHPBC}. The problem of finding a "good tetrad" is then transfered into
the problem to find an appropriate spin connections. In highly symmetric systems, like FRW cosmological background or
spherically symmetric black holes, symmetry considerations are enough to find corresponding spin connections. However,
situation is much less clear in more general systems.

One of the most important problem in cosmology (from both theoretical and observational perspectives) is to study evolution
of small perturbations over a cosmological background. From the viewpoint of good tetrad this problem have been considered in 
\cite{G1,G4}. As for the covariant approach, in the perturbed system
 we can not use any more symmetry properties of the FRW solution. 
It is not even known  if the problem of finding spin connection for a perturbed cosmological system
has a unique solution.  
In our recent paper \cite{we} we show that general properties of spin connections  allows us to find
at least one particular form of spin connection, leading to consistent equations of motion for linear scalar cosmological
perturbations on FRW background. In the present paper we generalize this result for more general class of
modified gravity theories, which include a non-minimally coupled scalar field \cite{GLSW,GLS,SST,JT,GEOVS, CCLS}. It is known that non-minimal coupling to torsion leads to a theory different from
the theory with the same coupling to curvature, so scalar-torsion theories represents one more class of modified gravity theories. In the present paper we assume that both
$f(T)$ modification and a scalar field with non-minimal coupling to torsion are present in the theory. Our considerations mostly follow those
of \cite{we}, so we mostly concentrate on differences between that paper and the present more general situation. Some 
basic issues related to spin connections are repeated to make the present paper self-consistent

We start with a brief outline of the properties of covariant formulation of 
teleparallel gravity. Within this approach we use the following definition of connections 
\begin{equation}
\Gamma^{\lambda}_{\,\,\,\,\mu\nu}=e_A^{\,\,\,\,\lambda}\left ( \partial_{\nu}e^A_{\,\,\,\,\mu} +\omega^{A}_{\,\,\,\,B\nu} e^B_{\,\,\,\,\mu}  \right),
\label{1.10}
\end{equation}
where $\omega^{A}_{\,\,\,\,B\nu}$ is so called spin connection.

One of the important consequence of this approach is possibility to restore local Lorentz invariance of  the theory. To do this the spin connection should
 satisfy  the set of important conditions. First of all
 it is antisymmetric with respect to Latin indexes:
\begin{equation}
\omega^{AB}_{\,\,\,\,\,\,\,\,\mu}=-\omega^{BA}_{\,\,\,\,\,\,\,\,\mu},
\label{1.11}
\end{equation}
where Latin indexes both must be in upper or lower position. Second,
spin connection are flat in the sense that
\begin{equation}
R^{A}_{\,\,\,\,B\mu\nu}(\omega^A_{\,\,\,\,B\mu})=\partial_{\mu}\omega^{A}_{\,\,\,\,B\nu} -\partial_{\nu}\omega^{A}_{\,\,\,\,B\mu}+\omega^{A}_{\,\,\,\,C\mu}\omega^{C}_{\,\,\,\,B\nu} -\omega^{A}_{\,\,\,\,C\nu}\omega^{C}_{\,\,\,\,B\mu}=0.
\label{1.12}
\end{equation}

All this philosophy assumes that tetrad, which correspond to our metric is not defined uniquely, but only up to a local Lorenz transformation, which transforms the spin connection as well \cite{HJU}
\begin{equation}
e'^A_{\,\,\,\,\mu}=\Lambda^A_{\,\,\,\,B} e^B_{\,\,\,\,\mu},\,\,\,\, \omega'^A_{\,\,\,\,\,\,\,\,B\mu}=\Lambda^A_{\,\,\,\,C}\omega^{C}_{\,\,\,\,F\mu}\Lambda_B^{\,\,\,\,F} +\Lambda^A_{\,\,\,\,C}\partial_{\mu}\Lambda_B^{\,\,\,\,C},
\label{1.13}
\end{equation}
where $\Lambda_A^{\,\,\,\,B}$ is the inverse of the Lorenz transformation matrix $\Lambda^A_{\,\,\,\,B}$. The third condition to be satisfied is the
 metricity condition
\begin{equation}
\partial_{\mu} e^A_{\,\,\,\,\nu}+ \omega^{A}_{\,\,\,\,B\mu} e^B_{\,\,\,\,\nu} -\Gamma^{\lambda}_{\,\,\,\,\nu\mu}e^A_{\,\,\,\,\lambda}=0.
\label{1.13.1}
\end{equation}

The definitions of torsion and contorsion are the same as in the old variant of the theory with zero spin connections,
$$
T^{\lambda}_{\,\,\,\,\mu\nu}\equiv \Gamma^{\lambda}_{\,\,\,\,\nu\mu} -\Gamma^{\lambda}_{\,\,\,\,\mu\nu},\,\,\,\,K^{\lambda}_{\,\,\,\,\mu\nu}\equiv   \Gamma^{\lambda}_{\,\,\,\,\mu\nu} -\stackrel{\!\!\!\!\!\!\!\!\!\!\!\circ}{\Gamma^{\lambda}_{\,\,\,\,\mu\nu}},
$$
where $\stackrel{\!\!\!\!\!\!\!\!\!\!\!\circ}{\Gamma^{\rho}_{\,\,\,\,\mu\nu}}=\frac{1}{2}g^{\rho\sigma}(\partial_{\mu}g_{\sigma\nu}+\partial_{\nu}g_{\mu\sigma} -\partial_{\sigma}g_{\mu\nu})$ is the Levi-Civita connection,
however, all tensors  depend on spin connection implicitly through (\ref{1.10}).

The torsion scalar can
 be written in the form
 \begin{equation}
T\equiv \frac{1}{2} S_{\lambda}^{\,\,\,\,\mu\nu}T^{\lambda}_{\,\,\,\,\mu\nu}  =\frac{1}{4}T^{\lambda\mu\nu}T_{\lambda\mu\nu}+\frac{1}{2}T^{\lambda\mu\nu}T_{\nu\mu\lambda}- T_{\lambda\mu}^{\,\,\,\,\,\,\,\,\lambda}T^{\nu\mu}_{\,\,\,\,\,\,\,\,\nu},
\label{1.6}
\end{equation}
where
\begin{equation}
S_{\lambda}^{\,\,\,\,\mu\nu}\equiv (K^{\mu\nu}_{\,\,\,\,\,\,\,\,\lambda}+\delta^{\mu}_{\lambda} T^{\alpha\nu}_{\,\,\,\,\,\,\,\,\alpha} -\delta^{\nu}_{\lambda} T^{\alpha\mu}_{\,\,\,\,\,\,\,\,\alpha})
\label{1.5}
\end{equation}
and the action of teleparallel gravity for the $f(T)$ theory can be written as
\begin{equation}
S=\frac{1}{2\kappa^2}\int d^4x\, e\, f(T),
\label{1.7}
\end{equation}
where $e=det(e^A_{\,\,\,\,\mu})=\sqrt{-g}$ and $\kappa^2$ is the gravitational constant. Variation of action (\ref{1.7}) with respect to tetrad field give us equation of motion in the following form
\begin{equation}
e^{-1}\partial_{\mu}(eS_A^{\,\,\,\,\nu\mu})f'-e^{\,\,\,\,\lambda}_AT^{\rho}_{\,\,\,\,\mu\lambda}S_{\rho}^{\,\,\,\,\mu\nu}f' + S_A^{\,\,\,\,\mu\nu}\partial_{\mu}(T)f''+\frac{1}{2}e^{\,\,\,\,\nu}_Af=\kappa^2 e^{\,\,\,\,\rho}_A \Theta_{\,\rho}^{\,\nu},
\label{1.8}
\end{equation}
which is also may be rewritten in equivalent form \cite{Barrow3}
\begin{equation}
\tilde{G}_{\mu\nu}\equiv f'(\stackrel{\!\!\!\!\!\!\circ}{R_{\mu\nu}}-\frac{1}{2}g_{\mu\nu}\stackrel{\circ}{R}) + \frac{1}{2}g_{\mu\nu}[f(T)-f'T] +f''S_{\nu\mu\lambda}\nabla^{\lambda}T=\kappa^2 \Theta_{\mu\nu},
\label{1.9}
\end{equation}
where $\stackrel{\!\!\!\!\!\!\circ}{G_{\mu\nu}}\equiv \stackrel{\!\!\!\!\!\!\circ}{R_{\mu\nu}}-\frac{1}{2}g_{\mu\nu}\stackrel{\circ}{R}$ should
be calculated  using Levi-Civita connection $\stackrel{\!\!\!\!\!\!\!\!\!\!\!\circ}{\Gamma^{\rho}_{\,\,\,\,\mu\nu}}$, and $\Theta_{\mu\nu}$ is the matter energy-momentum tensor.

The covariant  formulation of theory assumes that spin connection and tetrad are independent gravitational variables. So that, there is additional equation of motion, which appears due to  variation with respect to spin connection:

\begin{equation}
\partial_{\mu}f_T \left [ \partial_{\nu}(ee_{[A}^{\,\,\,\,\,\mu}e_{B]}^{\,\,\,\,\,\nu}) +2ee_C^{\,\,\,\,[\mu}e_{[A}^{\nu]}\omega^{C}_{\,\,\,\,B]\nu} \right ]=0,
\label{1.14}
\end{equation}
and this equation may be interpreted as equation for spin connection.

It was proved \cite{G3} that equation (\ref{1.14}) coincides  with non-symmetric part of equation of motion in the old formulation of teleparallel gravity, and therefore in the covariant formulation of teleparallel gravity the equation of motion is symmetric automatically.

In the present paper we consider more general teleparallel theory with the action 
\begin{equation}
S=\frac{1}{2\kappa^2}\int d^4x\, e\, [f(T,\Phi)+Z(\Phi)\partial^{\alpha}\Phi\partial_{\alpha}\Phi],
\label{1.15}
\end{equation}

The equations for tetrad and scalar field  are\cite{HJU}:
\begin{equation}
\bar{G}_{\mu\nu}\equiv f_T\stackrel{\!\!\!\!\!\!\circ}{G_{\mu\nu}} + \frac{1}{2}g_{\mu\nu}[f-f_TT] +S_{\nu\mu\lambda}\nabla^{\lambda}f_T - Z\nabla_{\mu}\Phi\nabla_{\nu}\Phi +\frac{1}{2}Zg_{\mu\nu}\nabla^{\alpha}\Phi\nabla_{\alpha}\Phi =\kappa^2 \Theta_{\mu\nu},
\label{1.16}
\end{equation}

\begin{equation}
f_{\Phi} -Z_{\Phi}\partial^{\alpha}\Phi\partial_{\alpha}\Phi - 2Z \stackrel{\circ}{\Box}\Phi = 0.
\label{1.18}
\end{equation}
As the spin connection enters the action of the theory only through the torsion scalar $T$, the equation for spin connection is the same as in $f(T)$ theory:
\begin{equation}
\partial_{\mu}f_T \left [ \partial_{\nu}(ee_{[A}^{\,\,\,\,\,\mu}e_{B]}^{\,\,\,\,\,\nu}) +2ee_C^{\,\,\,\,[\mu}e_{[A}^{\nu]}\omega^{C}_{\,\,\,\,B]\nu} \right ]=0,
\label{1.17}
\end{equation}
note, however, that now $\partial_{\mu}f_T = f_{TT} \partial_{\mu} T + f_{T \Phi} \partial_{\mu}\Phi$, so scalar field enters
into this equation implicitly.

To investigate the problem of cosmological perturbations we write down  the perturbed FRW-metric in the form
\begin{equation}
g_{\mu\nu}=diag\big [(1+2\psi), -a^2(1+2\phi), -a^2(1+2\phi), -a^2(1+2\phi)  \big ],
\label{2.1}
\end{equation}
and corresponding tetrad field
\begin{equation}
e^A_{\,\,\,\mu}=diag\big[ 1+\psi, a(1+\phi), a(1+\phi), a(1+\phi)   \big ],
\label{2.2}
\end{equation}
where $a=a(t)$ is scale factor and $\psi=\psi(t,x,y,z)$, $\phi=\phi(t,x,y,z)$ are small perturbations of metric. The inverse
metric and tetrad field matrix may be easily found as
$$
g^{\mu\nu}=diag\big [(1-2\psi), -a^{-2}(1-2\phi), -a^{-2}(1-2\phi), -a^{-2}(1-2\phi)  \big ],
$$
and
$$
e_A^{\,\,\,\mu}=diag\big[ 1-\psi, a^{-1}(1-\phi), a^{-1}(1-\phi), a^{-1}(1-\phi)   \big ].
$$

We have chosen the Cartesian tetrad since it is a proper tetrad for the unperturbed system. So that the spin connection can be set 
to zero without perturbations, and non-zero values of spin connection should be of the order of perturbations.

We start with antisymmetry condition (\ref{1.11}) for spin connection. It mean that in the most general case there are $24$ independent non-trivial components (all $\omega^A_{\,\,\,\,A\mu}=0$)
The independent $\omega^A_{\,\,\,\,B\mu}$ are:
$$
\begin{array}{c}
\omega^0_{\,\,\,10},\,\,\omega^0_{\,\,\,11},\,\,\omega^0_{\,\,\,12},\,\,\omega^0_{\,\,\,13},\,\, \omega^0_{\,\,\,20},\,\,\omega^0_{\,\,\,21},\,\,\omega^0_{\,\,\,22},\,\,\omega^0_{\,\,\,23},\,\,
\omega^0_{\,\,\,30},\,\,\omega^0_{\,\,\,31},\,\,\omega^0_{\,\,\,32},\,\,\omega^0_{\,\,\,33},\,\,\\
\\ \omega^1_{\,\,\,20},\,\,\omega^1_{\,\,\,21},\,\,\omega^1_{\,\,\,22},\,\,\omega^1_{\,\,\,23},\,\,
\omega^1_{\,\,\,30},\,\,\omega^1_{\,\,\,31},\,\,\omega^1_{\,\,\,32},\,\,\omega^1_{\,\,\,33},\,\,
\omega^2_{\,\,\,30},\,\,\omega^2_{\,\,\,31},\,\,\omega^2_{\,\,\,32},\,\,\omega^2_{\,\,\,33},\,\,
\end{array}
$$
and the dependent are
$$
\begin{array}{c}
\omega^1_{\,\,\,00}=\omega^0_{\,\,\,10},\,\,\omega^1_{\,\,\,01}=\omega^0_{\,\,\,11},\,\,\omega^1_{\,\,\,02}=\omega^0_{\,\,\,12}, \,\,\omega^1_{\,\,\,03}=\omega^0_{\,\,\,13},\,\, \omega^2_{\,\,\,00}=\omega^0_{\,\,\,20},\,\,\omega^2_{\,\,\,01}=\omega^0_{\,\,\,21},\,\,\\
\\ \omega^2_{\,\,\,02}=\omega^0_{\,\,\,22},\,\,\omega^2_{\,\,\,03}=\omega^0_{\,\,\,23},\,\,
\omega^3_{\,\,\,00}=\omega^0_{\,\,\,30},\,\,\omega^3_{\,\,\,01}=\omega^0_{\,\,\,31},\,\,\omega^3_{\,\,\,02}=\omega^0_{\,\,\,32}, \,\,\omega^3_{\,\,\,03}=\omega^0_{\,\,\,33},\,\,\\
\\ \omega^2_{\,\,\,10}=-\omega^1_{\,\,\,20},\,\,\omega^2_{\,\,\,11}=-\omega^1_{\,\,\,21},\,\,\omega^2_{\,\,\,12}=-\omega^1_{\,\,\,22}, \,\,\omega^2_{\,\,\,13}=-\omega^1_{\,\,\,23},\,\,
\omega^3_{\,\,\,10}=-\omega^1_{\,\,\,30},\,\,\omega^3_{\,\,\,11}=-\omega^1_{\,\,\,31},\,\,\\
\\ \omega^3_{\,\,\,12}=-\omega^1_{\,\,\,32},\,\,\omega^3_{\,\,\,13}=-\omega^1_{\,\,\,33},\,\,
\omega^3_{\,\,\,20}=-\omega^2_{\,\,\,30},\,\,\omega^3_{\,\,\,21}=-\omega^2_{\,\,\,31},\,\,\omega^3_{\,\,\,22}=-\omega^2_{\,\,\,32}, \,\,\omega^3_{\,\,\,23}=-\omega^2_{\,\,\,33},
\end{array}
$$
all this values depend on all coordinates in the most general case.

Now taking the tetrad (\ref{2.2}) and all $24$ independent components of $\omega^A_{\,\,\,\,B\mu}$ we can calculate all values which is needed for equation of motion (\ref{1.16}), keeping only linear terms with respect to $\psi$, $\phi$, $\omega^A_{\,\,\,\,B\mu}$ and their  derivatives.
As the 
expressions for $T^{\lambda}_{\,\,\,\,\mu\nu}$ and $S^{\lambda}_{\,\,\,\,\mu\nu}$ are rather combersome,  we write down here final results only for torsion scalar and for $S_{ikl}\partial^l f_T$ product because they are
both important for equations of motion.

Calculation for the torsion scalar give us the following result:
\begin{equation}
T=-6H^2+12H^2\psi -12H\dot\phi + 4\frac{H}{a}\big (\omega^0_{\,\,\,11}+\omega^0_{\,\,\,22}+\omega^0_{\,\,\,33}  \big),
\label{2.3}
\end{equation}
and  we  can  see  that  in  spite  of  the most  general  form  of  spin  connection  only  three  components  are  present  in  this
expression.
For the scalar field perturbations we take $\Phi=\Phi(t)+\delta\Phi(t,x,y,z)\equiv\Phi_0+\delta\Phi$.

Consider again the equation of motion (\ref{1.16}). Any non-symmetric term can arise from the only source $S_{\nu\mu\lambda}\nabla^{\lambda}f_T=f_{TT}S_{\nu\mu\lambda}\nabla^{\lambda}T+f_{T\Phi}S_{\nu\mu\lambda}\nabla^{\lambda}\Phi\equiv f_{TT}M_{\nu\mu}+f_{T\Phi}N_{\nu\mu}$. Calculating its components up to the first order in perturbations we have:
\begin{eqnarray}
M_{00}=&&0,
\label{2.4.1}\\
M_{01}=&& 12\frac{\dot a}{a^3}\big (a\ddot a- \dot a^2\big )\big ( \omega^1_{\,\,\,22} + \omega^1_{\,\,\,33} +2\phi_x \big ) ,
\label{2.4.2}\\
M_{02}=&& 12\frac{\dot a}{a^3}\big ( a\ddot a- \dot a^2\big )\big ( \omega^2_{\,\,\,33} - \omega^1_{\,\,\,21} +2\phi_y \big ),
\label{2.4.3}\\
M_{03}=&& 12\frac{\dot a}{a^3}\big ( a\ddot a- \dot a^2\big )\big ( -\omega^1_{\,\,\,31} - \omega^2_{\,\,\,32} +2\phi_z \big ),
\label{2.4.4}\\
M_{10}=&& -8\frac{\dot a^2}{a^3} \frac{\partial}{\partial x} \Big ( \omega^0_{\,\,\,11}+\omega^0_{\,\,\,22}+\omega^0_{\,\,\,33} - 3a\phi_t+3\dot a\psi \Big),
\label{2.4.5}\\
M_{11}=&& M_0+ M_* + 12\frac{H}{a} (a\ddot a- \dot a^2)\omega^0_{\,\,\,11}  ,
\label{2.4.6}\\
M_{12}=&& 12\frac{\dot a}{a^2}\big ( a\ddot a- \dot a^2\big ) \omega^0_{\,\,\,21},
\label{2.4.7}\\
M_{13}=&&  12\frac{\dot a}{a^2}\big ( a\ddot a- \dot a^2\big ) \omega^0_{\,\,\,31},
\label{2.4.8}\\
M_{20}=&& -8\frac{\dot a^2}{a^3} \frac{\partial}{\partial y} \Big ( \omega^0_{\,\,\,11}+\omega^0_{\,\,\,22}+\omega^0_{\,\,\,33} - 3a\phi_t+3\dot a\psi \Big),
\label{2.4.9}\\
M_{21}=&&  12\frac{\dot a}{a^2}\big ( a\ddot a- \dot a^2\big ) \omega^0_{\,\,\,12},
\label{2.4.10}\\
M_{22}=&& M_0+ M_*  + 12\frac{H}{a} (a\ddot a- \dot a^2)\omega^0_{\,\,\,22} ,
\label{2.4.11}\\
M_{23}=&&  12\frac{\dot a}{a^2}\big ( a\ddot a- \dot a^2\big ) \omega^0_{\,\,\,32},
\label{2.4.12}\\
M_{30}=&& -8\frac{\dot a^2}{a^3} \frac{\partial}{\partial z} \Big ( \omega^0_{\,\,\,11}+\omega^0_{\,\,\,22}+\omega^0_{\,\,\,33} - 3a\phi_t+3\dot a\psi \Big),
\label{2.4.13}\\
M_{31}=&&  12\frac{\dot a}{a^2}\big ( a\ddot a- \dot a^2\big ) \omega^0_{\,\,\,13},
\label{2.4.14}\\
M_{32}=&&  12\frac{\dot a}{a^2}\big ( a\ddot a- \dot a^2\big ) \omega^0_{\,\,\,23},
\label{2.4.15}\\
M_{33}=&& M_0+M_*+ 12\frac{H}{a} (a\ddot a- \dot a^2)\omega^0_{\,\,\,33}  ,
\label{2.4.16}
\end{eqnarray}
with
\begin{eqnarray}
M_0=&&-24\Big[ H^2(\dot a^2-a\ddot a)\Big] - 24\Big[ H^2a\dot a\dot\psi -4H^2(\dot a^2-a\ddot a)\psi +2H^2(\dot a^2-a\ddot a)\phi +2H(\dot a^2-a\ddot a)\dot\phi -\dot a^2\ddot\phi \Big]\label{2.4.17}\\
M_*=&&4\frac{H}{a}\Big[ (7\dot a^2-5a\ddot a)(\omega^0_{\,\,\,11}+\omega^0_{\,\,\,22}+\omega^0_{\,\,\,33}) -2a\dot a  \frac{\partial}{\partial t}(\omega^0_{\,\,\,11}+\omega^0_{\,\,\,22}+\omega^0_{\,\,\,33})   \Big],
\label{2.4.18}
\end{eqnarray}
and
\begin{eqnarray}
N_{00}=&&0,
\label{2.4.19}\\
N_{01}=&& -\dot\Phi_0( \omega^1_{\,\,\,22} + \omega^1_{\,\,\,33}+ 2\phi_x),
\label{2.4.20}\\
N_{02}=&& -\dot\Phi_0(\omega^2_{\,\,\,33} -\omega^1_{\,\,\,21} + 2\phi_y),
\label{2.4.21}\\
N_{03}=&&  -\dot\Phi_0( - \omega^1_{\,\,\,31} -\omega^2_{\,\,\,32} +2\phi_z),
\label{2.4.22}\\
N_{10}=&& -2H\delta\Phi_x,
\label{2.4.23}\\
N_{11}=&& N_0+ N_* -a\dot\Phi_0\omega^0_{\,\,\,11}  ,
\label{2.4.24}\\
N_{12}=&& -a\dot\Phi_0 \omega^0_{\,\,\,21},
\label{2.4.25}\\
N_{13}=&&  -a\dot\Phi_0 \omega^0_{\,\,\,31},
\label{2.4.26}\\
N_{20}=&& -2H\delta\Phi_y,
\label{2.4.27}\\
N_{21}=&&  -a\dot\Phi_0 \omega^0_{\,\,\,12},
\label{2.4.28}\\
N_{22}=&& N_0+ N_*  -a\dot\Phi_0\omega^0_{\,\,\,22} ,
\label{2.4.29}\\
N_{23}=&&  -a\dot\Phi_0 \omega^0_{\,\,\,32},
\label{2.4.30}\\
N_{30}=&& -2H\delta\Phi_z,
\label{2.4.31}\\
N_{31}=&&  -a\dot\Phi_0 \omega^0_{\,\,\,13},
\label{2.4.32}\\
N_{32}=&& -a\dot\Phi_0 \omega^0_{\,\,\,23},
\label{2.4.33}\\
N_{33}=&& N_0+N_*-a\dot\Phi_0\omega^0_{\,\,\,33}  ,
\label{2.4.34}
\end{eqnarray}
with
\begin{eqnarray}
N_0=&&-2a\dot a\dot\Phi_0 +4a\dot a\dot\Phi_0\psi-2a\dot a\delta\dot\Phi-2a\dot\Phi_0(a\dot\phi+2\dot a\phi), \label{2.4.35}\\
N_*=&&a\dot\Phi_0(\omega^0_{\,\,\,11}+\omega^0_{\,\,\,22}+\omega^0_{\,\,\,33}) .
\label{2.4.36}
\end{eqnarray}

We can see that for the non-symmetric part to vanish it is needed that
\begin{equation}
 f_{TT}M_{\nu\mu}+f_{T\Phi}N_{\nu\mu}= f_{TT}M_{\mu\nu}+f_{T\Phi}N_{\mu\nu},
\label{2.4.36.1}
\end{equation}
so we are forced to put
$$
\omega^0_{\,\,\,ij}=\omega^0_{\,\,\,ji},
$$
and
\begin{eqnarray}
\big(12H\dot Hf^0_{TT}-\dot\Phi_0f^0_{T\Phi} \big)\big( \omega^1_{\,\,\,22}+\omega^1_{\,\,\,33} \big)=&& 2\partial_x\left[ 4Hf^0_{TT} \big( 3H\dot\phi -3\dot H\phi -3H^2\psi -\frac{H}{a}\omega^0 \big) +f^0_{T\Phi}\big( \dot\Phi_0\phi- H\delta\Phi \big) \right],
\label{2.4.37}\\
\big( 12H\dot Hf^0_{TT}-\dot\Phi_0f^0_{T\Phi} \big)\big(\omega^2_{\,\,\,33}-\omega^1_{\,\,\,21} \big)=&& 2\partial_y\left[ 4Hf^0_{TT} \big( 3H\dot\phi -3\dot H\phi -3H^2\psi -\frac{H}{a}\omega^0 \big) +f^0_{T\Phi}\big( \dot\Phi_0\phi- H\delta\Phi \big) \right],
\label{2.4.38}\\
-\big(12H\dot Hf^0_{TT}-\dot\Phi_0f^0_{T\Phi} \big)\big( \omega^1_{\,\,\,31}+\omega^2_{\,\,\,32} \big)=&& 2\partial_z\left[ 4Hf^0_{TT} \big( 3H\dot\phi -3\dot H\phi -3H^2\psi -\frac{H}{a}\omega^0 \big) +f^0_{T\Phi}\big( \dot\Phi_0\phi- H\delta\Phi \big) \right],
\label{2.4.39}
\end{eqnarray}
where we denote $\omega^0\equiv\omega^0_{\,\,\,11}+\omega^0_{\,\,\,22}+\omega^0_{\,\,\,33}$ and $f^0_{TT}\equiv f_{TT}(T_0,\Phi_0)$, $f^0_{T\Phi}\equiv f_{T\Phi}(T_0,\Phi_0)$.

The general form of the flatness condition is the same as in \cite{we}, so we briefly repeat the corresponding expressions.
 We can neglect two last terms in (\ref{1.12}), because they are
second order in perturbations.  Thus the conditions can be presented in the following form
$$
\partial_{\mu}\omega^A_{\,\,\,B\nu}=\partial_{\nu}\omega^A_{\,\,\,B\mu},
$$
which implies that
\begin{eqnarray}
&&\omega^0_{\,\,\,10}= a_{1t},\,\,\,\omega^0_{\,\,\,11}= a_{1x},\,\,\,\omega^0_{\,\,\,12}= a_{1y},\,\,\,\omega^0_{\,\,\,13}= a_{1z},
\label{2.4.40}\\
&&\omega^0_{\,\,\,20}= a_{2t},\,\,\,\omega^0_{\,\,\,21}= a_{2x},\,\,\,\omega^0_{\,\,\,22}= a_{2y},\,\,\,\omega^0_{\,\,\,23}= a_{2z},
\label{2.4.41}\\
&&\omega^0_{\,\,\,30}= a_{3t},\,\,\,\omega^0_{\,\,\,31}= a_{3x},\,\,\,\omega^0_{\,\,\,32}= a_{3y},\,\,\,\omega^0_{\,\,\,33}= a_{3z},
\label{2.4.42}\\
&&\omega^1_{\,\,\,20}=  b_{1t},\,\,\,\omega^1_{\,\,\,21}= b_{1x},\,\,\,\omega^1_{\,\,\,22}= b_{1y},\,\,\,\omega^1_{\,\,\,23}= b_{1z},
\label{2.4.43}\\
&&\omega^1_{\,\,\,30}= b_{2t},\,\,\,\omega^1_{\,\,\,31}= b_{2x},\,\,\,\omega^1_{\,\,\,32}= b_{2y},\,\,\,\omega^1_{\,\,\,33}= b_{2z},
\label{2.4.44}\\
&&\omega^2_{\,\,\,30}=  c_{t},\,\,\,\,\,\omega^2_{\,\,\,31}= c_{x},\,\,\,\,\,\omega^2_{\,\,\,32}= c_{y},\,\,\,\,\,\omega^2_{\,\,\,33}= c_{z},
\label{2.4.45}
\end{eqnarray}
where $a_i$, $b_i$ and $c$ are some functions of the coordinates $(t,x,y,z)$ and index $\mu=t,x,y,z$ denotes differentiation with respect to coordinate.

We now start to search for a solution for spin connections.
First of all we write down $\{ij\}$-components of equation (\ref{1.16}) for $i\neq j$:
\begin{equation}
f^0_T\stackrel{\!\!\!\!\!\!\circ}{R_{ij}}+f^0_{TT}M_{ij}+f^0_{T\Phi}N_{ij} \underset{i\neq j}{=} -f^0_T\partial_i\partial_j (\phi+\psi)+ \left(12f^0_{TT}H\dot Ha -f^0_{T\Phi}a\dot\Phi_0 \right)\omega^0_{\,\,\,ij}=\kappa^2\pi_{ij}.
\label{3.1}
\end{equation}
Implying zero anisotropic stress and using the  parametrization (\ref{2.4.40})-(\ref{2.4.42}), which tell us that $\omega^0_{\,\,\,ij}=\omega^0_{\,\,\,ji}=\partial_ja_i$, we rewrite equation (\ref{3.1}) as follows
$$
f^0_T\partial_i\partial_j (\phi+\psi)= \left(12f^0_{TT}H\dot Ha -f^0_{T\Phi}a\dot\Phi_0 \right)\partial_ja_i,
$$
which after integration gives us a general expression for $a_i$ functions:
\begin{equation}
a_i=A_0\partial_i (\phi+\psi),
\label{3.2}
\end{equation}
where we denote
\begin{equation}
A_0\equiv\frac{f^0_T}{12f^0_{TT}H\dot Ha -f^0_{T\Phi}a\dot\Phi_0}.
\label{3.3}
\end{equation}
Solution (\ref{3.2}) allows us to find a half of the components of spin connection set:
\begin{equation}
\omega^0_{\,\,\,i\mu}=A_0\partial_i\partial_{\mu} (\phi+\psi).
\label{3.4}
\end{equation}
In particular, we have
\begin{equation}
\omega^0=A_0\triangle(\phi+\psi).
\label{3.4.1}
\end{equation}

To write down the remaining half of components of spin connection let us go back to the system (\ref{2.4.37})-(\ref{2.4.39}). As in the pure $f(T)$ case \cite{we}, 
it can be rewritten in the form
$$
\mathbf{rot} \overrightarrow{B}=2\mathbf{grad}\Psi,
$$
where 3d vectors depend now on the scalar field perturbations as well:
 $\overrightarrow{B}$ is the 3d vector with the components $\overrightarrow{B}=\big(12H\dot Hf^0_{TT}-\dot\Phi_0f^0_{T\Phi} \big)(c,\,\,-b_2,\,\,b_1)$ and $\Psi = 4Hf^0_{TT} \big( 3H\dot\phi -3\dot H\phi -3H^2\psi -\frac{H}{a}\omega^0 \big) +f^0_{T\Phi}\big( \dot\Phi_0\phi- H\delta\Phi \big)$. After taking the divergence from left and right sides of the previous equation we find the following relation for perturbations
\begin{equation}
\triangle\left( 4Hf^0_{TT} \big( 3H\dot\phi -3\dot H\phi -3H^2\psi -\frac{H}{a}\omega^0 \big) +f^0_{T\Phi}\big( \dot\Phi_0\phi- H\delta\Phi \big)\right)=0,
\label{3.5}
\end{equation}
this expression together with (\ref{3.4.1}) can be used to determine the relation between $\phi$ and $\psi$, known as the gravitational slip. The simplest solution of this equation is
\begin{equation}
 4Hf^0_{TT} \big( 3H\dot\phi -3\dot H\phi -3H^2\psi -\frac{H}{a}\omega^0 \big) +f^0_{T\Phi}\big( \dot\Phi_0\phi- H\delta\Phi \big)=0,
\label{3.5.1}
\end{equation}
and therefore we have
$$
\mathbf{rot} \overrightarrow{B}=0.
$$
This indicates that the vector $\overrightarrow{B}$ is a gradient of some function $\alpha$,
and the solution for remaining components of spin connection is
\begin{equation}
c=\partial_x\alpha,\,\,\,b_2=-\partial_y\alpha,\,\,\,b_1=\partial_z\alpha,
\label{3.5.2}
\end{equation}
where $\alpha$ is some function of the order of perturbations, its
 explicit form 
  is not needed for closing the system of equations for perturbations.

Having the spin connection coefficients we can now turn to equations of motion.
First of all note that we write equations for perturbations  using (\ref{1.9}) with one upper and one lower indexes, which is more conventional  for this task.

\begin{equation}
\bar{G}^{\mu}_{\nu}\equiv f_T\stackrel{\!\!\!\circ}{G^{\mu}_{\nu}} + \frac{1}{2}\delta^{\mu}_{\nu}[f(T)-f_TT] +f_{TT}M^{\mu}_{\nu}+f_{T\Phi}N^{\mu}_{\nu}-Z\nabla^{\mu}\Phi\nabla_{\nu}\Phi+\frac{1}{2}Z\delta^{\mu}_{\nu}\nabla^{\alpha}\Phi\nabla_{\alpha}\Phi=\kappa^2 \Theta^{\mu}_{\nu}.
\label{3.6}
\end{equation}

Energy-momentum tensor has the following form
\begin{equation}
\Theta^{\mu}_{\nu}= (\rho+p)u^{\mu}u_{\nu}-\delta^{\mu}_{\nu}p,
\label{3.7}
\end{equation}
and
\begin{eqnarray}
\Theta^0_0&&= \delta\rho, \label{3.8}\\
\nonumber\\
\Theta^0_i&&= -(p+\rho)\partial_i\delta u, \label{3.9}\\
\nonumber\\
\Theta^i_j&&= -\delta^i_j\delta p, \label{3.10}
\end{eqnarray}

\vspace{1cm}

Non-trivial components of Einstein tensor  $
\stackrel{\!\!\!\circ}{G^{\mu}_{\nu}}\equiv\stackrel{\!\!\!\circ}{R^{\mu}_{\nu}}-\frac{1}{2}\delta^{\mu}_{\nu}\stackrel{\circ}{R}
$ read
\begin{equation}
\stackrel{\!\!\!\circ}{G^{0}_{0}}=3H^2+ 2 \left ( 3H\dot\phi -3H^2\psi -\frac{1}{a^2}\triangle \phi  \right ),
\label{3.11}
\end{equation}

\begin{equation}
\stackrel{\!\!\!\circ}{G^{0}_{i}}= -2\partial_i\left( \dot\phi - H\psi  \right ),
\label{3.12}
\end{equation}

\begin{equation}
\stackrel{\!\!\!\circ}{G^{i}_{i}}=2\dot H +3H^2 +2\left ( \ddot\phi +3H\dot\phi -H\dot\psi - 3H^2\psi - 2\dot H\psi   \right ) -\frac{1}{a^2}\left [\triangle(\phi+\psi)-\partial_i\partial_i(\phi+\psi) \right],
\label{3.13}
\end{equation}

To derive equations we need the following components of $M^{\mu}_{\nu}$:
\begin{equation}
M^0_{i} =  24H^2 \partial_i\big ( \dot\phi -H\psi -\frac{1}{3a}\omega^0  \big ) ,
\label{3.14}
\end{equation}

\begin{equation}
M^i_i=a^{-2}(2\phi-1)M_0 - a^{-2}( -8H\dot a\dot\omega^0 +28 H^2\dot a\omega^0 - 20H\ddot a\omega^0 +12\frac{H}{a}(a\ddot a-\dot a^2)\omega^0_{\,\,\,ii}),
\label{3.15}
\end{equation}
and of $N^{\mu}_{\nu}$:

\begin{equation}
N^0_{i} = -2H\partial_i\delta\Phi,
\label{3.16}
\end{equation}

\begin{equation}
N^i_i=2H\dot\Phi_0+2\dot\Phi_0\dot\phi-4H\dot\Phi_0\psi+2H\delta\dot\Phi-\frac{1}{a}\dot\Phi_0(\omega^0-\omega^0_{\,\,\,ii}).
\label{3.17}
\end{equation}

Note that we still have auxiliary variable in expressions, which can be excluded by using the relation (\ref{3.5.1}).  Together with (\ref{3.4.1}) it  produces the relation between $\phi$ and $\psi$ perturbations, 
determing the gravitational slip. As for the expression for $\omega^0_{\,\,\,ii}$, it be found by using (\ref{3.4}). After such kind of manipulations and simplifications we have instead of (\ref{3.15})
\begin{eqnarray}
M^i_i=&&-12H\dot H \frac{f^0_T}{12f^0_{TT}H\dot H -f^0_{T\Phi}\dot\Phi_0}\frac{1}{a^2}\partial_i\partial_{i} (\phi+\psi)-12\left(3\dot H^2\phi+2H\ddot H\phi+H\dot H\dot\phi-H^2\dot H\psi\right)\nonumber\\
\nonumber\\
&&+\frac{f^0_{T\Phi}}{f^0_{TT}}\left(2\dot\Phi_0\dot\phi+2\ddot\Phi_0\phi+\frac{\dot H}{H}\dot\Phi_0\phi -2H\delta\dot\Phi-3\dot H\delta\Phi \right) +2\left (\dot\Phi_0\phi-H\delta\Phi  \right )\frac{1}{f^0_{TT}} \left ( -12H\dot Hf^0_{TT\Phi}+f^0_{T\Phi\Phi}\dot\Phi_0  \right )\nonumber\\
\nonumber\\
&&-2\frac{f^0_{T\Phi}}{(f^0_{TT})^2}\left ( \dot\Phi_0\phi-H\delta\Phi \right )\left (-12H\dot Hf^0_{TTT}+f^0_{TT\Phi}\dot\Phi_0   \right )-24H^2\dot H,\nonumber
\end{eqnarray}
and instead of (\ref{3.17})
\begin{eqnarray}
N^i_i=2H\dot\Phi_0-\dot\Phi_0\dot\phi-H\dot\Phi_0\psi+2H\delta\dot\Phi+3\dot\Phi_0\frac{\dot H}{H}\phi - \frac{1}{3H^2}\dot\Phi_0\frac{f^0_{T\Phi}}{f^0_{TT}}\left ( \dot\Phi_0\phi-H\delta\Phi \right ) +\dot\Phi_0 \frac{f^0_T}{12f^0_{TT}H\dot H -f^0_{T\Phi}\dot\Phi_0}\frac{1}{a^2}\partial_i\partial_{i} (\phi+\psi).\nonumber
\end{eqnarray}

Finally decomposing derivatives of function $f$ and $Z$ in Taylor series near the point $T=T_0\equiv -6H^2$, $\Phi=\Phi_0$ we have

$$
f=f^0+f^0_T\delta T+f^0_{\Phi}\delta\Phi,
$$

$$
f_T=f^0_T+f^0_{TT}\delta T+f^0_{T\Phi}\delta\Phi,
$$

$$
f_{TT}=f^0_{TT}+f^0_{TTT}\delta T+f^0_{TT\Phi}\delta\Phi,
$$

$$
f_{T\Phi}=f^0_{T\Phi}+f^0_{TT\Phi}\delta T+f^0_{T\Phi\Phi}\delta\Phi,
$$

$$
Z=Z^0+Z^0_{\Phi}\delta\Phi,
$$

where  $f^0_T=f_T|_{T=T_0,\Phi=\Phi_0}$ etc. Remembering that  according to (\ref{2.3}) $\delta T= 12H^2\psi -12H\dot\phi+4\frac{H}{a}A_0\triangle(\phi+\psi)$,
we can collect all terms and get

\begin{eqnarray}
\bar{G}^0_0=&&    2f^0_T \left (3H\dot\phi- 3H^2\psi  -\frac{1}{a^2}\triangle \phi  \right ) - 72H^2\dot Hf^0_{TT}\phi+6H\dot\Phi_0f^0_{T\Phi}\phi +\frac{1}{2}f^0_{\Phi}\delta\Phi+\sigma^0_0= \kappa^2\delta\rho, \label{3.18}\\
\nonumber\\
\bar{G}^0_{i}=&& -2f^0_T\partial_i(\dot\phi-H\psi)+24f^0_{TT}H\dot H\partial_i\phi-2\dot\Phi_0f^0_{T\Phi}\partial_i\phi+\sigma^0_i= -\kappa^2(p+\rho)\partial_i\delta u, \label{3.19}\\
\nonumber\\
\bar{G}^i_i =&&  2 f^0_T \left (  \ddot\phi +3H\dot\phi -H\dot\psi - 3H^2\psi - 2\dot H\psi   -\frac{1}{2a^2}\triangle(\phi+\psi) \right) +\frac{1}{2}f^0_{\Phi}\delta\Phi \label{3.20}\\&&+288H^2\dot H^2f^0_{TTT} \phi  -12Hf^0_{TT} \left( H\dot H\dot\phi +2\ddot HH\phi  +5\dot H^2\phi +6H^2\dot H\phi -H^2\dot H\psi \right )\nonumber\\&& +f^0_{T\Phi}\left( 6\frac{\dot H}{H}\dot\Phi_0\phi +2\ddot\Phi_0\phi +6H\dot\Phi_0\phi +\dot\Phi_0\dot\phi-H\dot\Phi_0\psi -3\dot H\delta\Phi \right) - \frac{\dot\Phi_0}{3H^2}\frac{(f^0_{T\Phi})^2}{f^0_{TT}}\left( \dot\Phi_0\phi-H\delta\Phi \right)\nonumber\\
\nonumber\\
&&-48H\dot Hf^0_{TT\Phi}\dot\Phi_0\phi+2f^0_{T\Phi\Phi}\dot\Phi_0^2\phi+\sigma^i_i= -\kappa^2\delta p. \nonumber
\end{eqnarray}

where we denote
\begin{equation}
\sigma_{\mu\nu}\equiv - Z\nabla_{\mu}\Phi\nabla_{\nu}\Phi +\frac{1}{2}Zg_{\mu\nu}\nabla^{\alpha}\Phi\nabla_{\alpha}\Phi,
\label{3.20.1}
\end{equation}

so that
\begin{equation}
\sigma^0_0=-\frac{1}{2}\left( 2Z^0\dot\Phi_0\delta\dot\Phi-2Z^0\dot\Phi_0^2\psi+Z^0_{\Phi}\dot\Phi_0^2\delta\Phi \right),\,\,\,\,\,\, \sigma^0_i=-Z^0\dot\Phi_0\delta\Phi,\,\,\,\,\,\,\sigma^i_i=-\sigma^0_0.
\label{3.21}
\end{equation}
Finally, the field equation (\ref{1.18}) gives us

\begin{eqnarray}
&&-12\dot Hf^0_{T\Phi}\phi+\frac{(f^0_{T\Phi})^2}{Hf^0_{TT}}(\dot\Phi_0\phi-H\delta\Phi)+f^0_{\Phi\Phi}\delta\Phi + 2Z^0_{\Phi}\dot\Phi_0^2\psi-2Z^0_{\Phi}\dot\Phi_0\delta\dot\Phi-Z^0_{\Phi\Phi}\dot\Phi_0^2\delta\Phi -2Z^0_{\Phi}(\ddot\Phi_0+3H\dot\Phi_0)\delta\Phi\nonumber\\
\nonumber\\
&&-2Z^0\left[ (\delta\ddot\Phi+3H\delta\dot\Phi)-2\psi(\ddot\Phi_0+3H\dot\Phi_0) -\dot\Phi_0\dot\psi+3\dot\Phi_0\dot\phi+\triangle\delta\Phi \right]=0.
\label{3.22}
\end{eqnarray}

As for explicit equation for the gravitational slip,
it can be found by using two different representations (\ref{3.4.1}) and (\ref{3.5.1}) for $\omega^0$:
\begin{equation}
\frac{4Hf^0_T}{12H\dot Hf^0_{TT}-f^0_{T\Phi}\dot\Phi_0}\frac{1}{a^2}\triangle(\phi+\psi)= 12H\dot\phi-12\dot H\phi-12H^2\psi +\frac{f^0_{T\Phi}}{Hf^0_{TT}}(\dot\Phi_0\phi-H\delta\Phi).
\label{3.23}
\end{equation}
In the most general case we have five variables $\phi$, $\psi$, $\delta\rho$, $\delta u$ and $\delta\Phi$. To find them  have four equations  (\ref{3.18})-(\ref{3.20}), (\ref{3.22}) and the slip relation (\ref{3.23}). 

Note, that an attempt to find a solution with a 
 separate symmetry $M_{\mu\nu}=M_{\nu\mu}$ and $N_{\mu\nu}=N_{\nu\mu}$ would lead to an additional constraint on the value of
 a scalar field perturbation. Indeed, 
 these conditions written for  $N_{0i}$ components shows that in this case we have  $H\delta\Phi=\dot\Phi_0\phi$, making the whole
 system of equations overdetermined.

At that point we should note that resulting set of equations has no smooth limit for $f_{TT} \to 0$, as we see that  $f_{TT}$ enters in  denominators
  in the equations (\ref{3.20}), (\ref{3.22}) and (\ref{3.23}). This means that the case of a pure non-minimal coupling to torsion is still out of our considerations. To obtain equations for this special case we are forced to  start from  the very beginning. 

Without loss of generality we can put $f\equiv T F(\Phi)$ and field equations (\ref{1.16}), (\ref{1.18}) take the form
\begin{equation}
\tilde{G}_{\mu\nu}\equiv F\stackrel{\!\!\!\!\!\!\circ}{G_{\mu\nu}}  +F_{\Phi}N_{\nu\mu}  +\sigma_{\mu\nu}=\kappa^2 \Theta_{\mu\nu},
\label{4.1}
\end{equation}

\begin{equation}
TF_{\Phi} -Z_{\Phi}\partial^{\alpha}\Phi\partial_{\alpha}\Phi - 2Z \stackrel{\circ}{\Box}\Phi = 0,
\label{4.2}
\end{equation}

First of all we can see that symmetry conditions,  instead of (\ref{2.4.36.1}), are now
\begin{equation}
N_{\mu\nu}=N_{\nu\mu},
\label{4.3}
\end{equation}
moreover term $M_{\mu\nu}$ is totally disconnected from other equations and we do not need  it any more. Nevertheless, the condition 

$$
\omega^0_{\,\,\,ij}=\omega^0_{\,\,\,ji},
$$
 still holds. Now from $\{ij\}$-components of equation (\ref{1.16}) for $i\neq j$ we have the following solutions

\begin{equation}
\omega^0_{\,\,\,i\mu}=A_0\partial_i\partial_{\mu} (\phi+\psi),
\label{4.4}
\end{equation}
and
\begin{equation}
\omega^0=A_0\triangle(\phi+\psi)
\label{4.5}
\end{equation}
where, instead of (\ref{3.3}), 
\begin{equation}
A_0\equiv-\frac{f^0_T}{f^0_{T\Phi}a\dot\Phi_0}=-\frac{F^0}{F^0_{\Phi}a\dot\Phi_0}.
\label{4.6}
\end{equation}

Solution of equation (\ref{3.5}) takes in this case a very simple form
\begin{equation}
H\delta\Phi=\dot\Phi_0\phi,
\label{4.7}
\end{equation}
whereas solution for spatial part of spin connection (\ref{3.5.2}) is unchanged.

Now decomposing equation (\ref{4.1}) and using our previous calculations for $\stackrel{\!\!\!\circ}{G^{\mu}_{\nu}}$ and $N^{\mu}_{\nu}$ we obtain

\begin{eqnarray}
\tilde{G}^0_0=&&    2F^0 \left (3H\dot\phi- 3H^2\psi  -\frac{1}{a^2}\triangle \phi  \right ) +3H^2F^0_{\Phi}\delta\Phi+\sigma^0_0= \kappa^2\delta\rho, \label{4.8}\\
\nonumber\\
\tilde{G}^0_{i}=&& -2F^0\partial_i(\dot\phi-H\psi)-2HF^0_{\Phi}\partial_i\delta\Phi+\sigma^0_i= -\kappa^2(p+\rho)\partial_i\delta u, \label{4.9}\\
\nonumber\\
\tilde{G}^i_i =&&  2 F^0 \left (  \ddot\phi +3H\dot\phi -H\dot\psi - 3H^2\psi - 2\dot H\psi   \right) \label{4.10}\\&& +F^0_{\Phi} \left( 2\dot H\delta\Phi +3H^2\delta\Phi + 2\dot\Phi_0\dot\phi - 4H\dot\Phi_0\psi + 2H\delta\dot\Phi  \right ) + 2H\dot\Phi_0 F^0_{\Phi\Phi}\delta\Phi+\sigma^i_i= -\kappa^2\delta p, \nonumber
\end{eqnarray}
with $\sigma^{\mu}_{\nu}$ (\ref{3.21}).

Finally for field equation (\ref{4.2}) we have
\begin{eqnarray}
&& 12HF^0_{\Phi}(H\psi-\dot\phi)   -\frac{4H}{\dot\Phi_0}\frac{1}{a^2}\triangle(\phi+\psi)F^0           -6H^2F^0_{\Phi\Phi}\delta\Phi + 2Z^0_{\Phi}\dot\Phi_0^2\psi-2Z^0_{\Phi}\dot\Phi_0\delta\dot\Phi-Z^0_{\Phi\Phi}\dot\Phi_0^2\delta\Phi \nonumber\\
\nonumber\\
&&-2Z^0_{\Phi}(\ddot\Phi_0+3H\dot\Phi_0)\delta\Phi-2Z^0\left[ (\delta\ddot\Phi+3H\delta\dot\Phi)-2\psi(\ddot\Phi_0+3H\dot\Phi_0) -\dot\Phi_0\dot\psi+3\dot\Phi_0\dot\phi+\triangle\delta\Phi \right]=0.
\label{4.11}
\end{eqnarray}

Thus we can see that we have five equations (\ref{4.7})-(\ref{4.11}) for five variables: $\phi$, $\psi$, $\Phi$, $\delta\rho$, $\delta u$. Note, that in the case of $f_{TT}=0$ fixing $\delta \Phi$ using (\ref{4.7}) does not lead to any problems since now we have no analog of the slip equation -- the non-vanishing part of (71) does not contain $\omega_0$ which is connected with slip through (\ref{4.5}).
Nevertheless, we should treat two scalar potentials $\phi$ and $\psi$ as an independent variables, otherwise the number of equations will be bigger than the number of unknown variables, so the slip is
nontrivial for this special case as well.

In the present paper we have shown that the method of getting a consistent system of equations of motion for scalar linear cosmological perturbations
in $f(T)$ theory can be easily generalized for a wider class of scalar-torsion theories where function $f$ depends on torsion scalar $T$ and a scalar field $\Phi$. As in the case of $f(T)$ theory we do not consider the problem of 
uniqueness of spin connections, presenting instead a particular example of spin connections, leading to consistent equations of motion. We also have found that the general
form of equations of motion can not be applied to the case of usual non-minimal coupling between scalar field and torsion, i.e. when $f_{TT}(T, \Phi)$=0. This case needs a separate
treatment which is presented in our paper as well. We have written down equations of motion for this particular case, and note that they still require nontrivial gravitational slip, as  in general $f(T, \Phi)$ case,
though the resulting set of equations is a bit different.

\begin{acknowledgments}

The work of AT was supported by the Russian Science Foundation (RSF)
grant 21-12-00130
and by the Kazan Federal University Strategic Academic Leadership Program. The work of PT was supported by the RFBR grant 20-02-00411 A.

\end{acknowledgments}

\end{document}